\documentclass[mathematics,article,accept,pdftex,moreauthors]{Definitions/mdpi} 
\usepackage{extarrows}
\usepackage{amssymb}
\usepackage{multirow}
\usepackage{booktabs}
\usepackage{graphicx}
\usepackage[ruled,linesnumbered]{algorithm2e}

\newcommand{\highlight}[1]{\colorbox{green}{#1}}
\firstpage{1} 
\pubvolume{1}
\issuenum{1}
\articlenumber{0}
\pubyear{2023}
\copyrightyear{2023}
\externaleditor{Academic Editor: Carsten Schneider, Jonathan Blackledge}
\datereceived{20 May 2023} 
\daterevised{27 June 2023} 
\dateaccepted{5 July 2023} 
\datepublished{ } 
\hreflink{https://doi.org/} 

\Title{A Remote Quantum Error-Correcting Code Preparation Protocol on Cluster States}

\TitleCitation{A Remote Quantum Error-Correcting Code Preparation Protocol on Cluster States}

\Author{Qiang Zhao 
 $^{1,2}$, Haokun Mao $^{1}$, Yucheng Qiao $^{3}$, Ahmed A. Abd El-Latif $^{1,4,5}$ and Qiong Li $^{1,}$*}

\AuthorNames{Qiang Zhao, Haokun Mao, Yucheng Qiao, Ahmed A. Abd El-Latif and Qiong Li}

\AuthorCitation{Qiang Zhao
; Haokun Mao; Yucheng Qiao, ; A.A.A. El-Latif; Qiong Li.}

\address{%
$^{1}$ \quad Information Countermeasure Technique Institute, School of Cyberspace Science, Faculty of Computing, Harbin Institute of Technology, Harbin 150001
, China; {qiangzhao@cuhk.edu.hk} 
 (Q.Z.); \mbox{hkmao@hit.edu.cn 
 (H.M.)} \\ 
$^{2}$ \quad Department of Computer Science and Engineering, The Chinese University of Hong Kong, {Hong Kong 999077}, China\\
$^{3}$ \quad School of Information Engineering, Guilin University of Electronic Science and Technology, Guilin 541004, China; {glqyc251@guet.edu.cn}
 \\
$^{4}$ \quad EIAS Data Science Laboratory
, College of Computer and Information Sciences, Prince Sultan University, \linebreak Riyadh 11586, Saudi Arabia; {aabdellatif@psu.edu.sa}\\
$^{5}$ \quad Department of Mathematics and Computer Science, Faculty of Science, Menoufia University, \mbox{Shibin Al Kawm  32511}
, Egypt}
\corres{Correspondence: qiongli@hit.edu.cn}

\abstract{The blind quantum computation (BQC) protocol allows for privacy-preserving remote quantum computations. In this paper, we introduce a remote quantum error correction code preparation protocol for BQC using a cluster state and analyze its blindness in the measurement-based quantum computation model. Our protocol requires fewer quantum resources than previous methods, as it only needs weak coherent pulses, eliminating the need for quantum memory and limited quantum computing. The results of our theoretical analysis and simulations show that our protocol requires fewer quantum resources compared to non-coding methods with the same qubit error rate.}

\keyword{blind quantum computation; quantum error-correcting code; cluster state; preparation} 

\MSC{81P73
}
\begin{document}

\section{Introduction}
\label{sec:intro}
With the rapid development of quantum technology, quantum computing has attracted increasing attention from researchers because of its theoretically super-high computing power. In the classical field, it is very hard for conventional computers to deal with the non-deterministic polynomial (NP) problems. However, these tasks have the potential to be solved by a quantum computer in the future \cite{grover1996afast,shor1999polynomial,tanaka2017quantum,albash2018adiabatic,alsubai2023quantum,Lee2023depth}. At present, a practical quantum computer requires large-scale quantum resources, expensive equipment, and low-temperature environment, etc., which ordinary clients cannot afford. A blind quantum computation (BQC) protocol as an optimal potential solution can address this problem at hand. By harnessing of the power of BQC, the client (Alice) can effectively delegate intricate computing tasks to a remote quantum server (Bob) while guaranteeing the privacy of Alice's information \cite{childs2005secure,arrighi2006blind,broadbent2009universal}. The breakthrough of QBC is not only the low cost of overcoming Alice's computing limitations but also protecting sensitive data well. In recent years, many extension BQC protocols have emerged~\cite{dunjko2012blind,Xu2015Blind,Zhao2017Blind,Zhao2018Finite,Zhao2020fault,kashefi2021classical,shan2021multi}. The universal blind quantum computation (UBQC) proposed by Broadbent, Fitzsimons, and Kashefi~\cite{broadbent2009universal} is very popular among these protocols, in which Alice only needs to prepare the single-photon states.

In UBQC, the preparation is a very important process~\cite{broadbent2009universal} which directly determines the number of pulses to be sent and the state of the desired qubits and is the basis of the successful execution of subsequent quantum computing. In Alice's preparation, she prepares the required single photon pulses for a desired computing and sends them to Bob through a quantum channel. However, the probability of the photon number satisfies the Poisson distribution, which results in a lower generated probability of single photons. It is also inevitable that two or multiple photon pulses are sent in the preparation, which will destroy the privacy of quantum states in Alice's preparation. Hence, a remote blind qubit state preparation (RBSP) protocol is proposed by Dunjko et al.~\cite{dunjko2012blind} to delegate Bob to prepare qubits, which only requires Alice to send weak coherent pulses to Bob. To prepare a desired qubit, the number of pulses is of the order $O(1/T^{4})$, where $T(T<1)$ represents the transmittance of the quantum channel. Xu and Lo then presented the one decoy state-based RBSP protocol to improve the preparation efficiency~\cite{Xu2015Blind}. The number of pulses to prepare a desired qubit can be reduced from $O(1/T^{4})$ to almost $O(1/T)$. Obviously, the decoy state technique can be used to prepare more qubits in unit time. On this basis, the multi-decoy states technique in quantum key distribution~\cite{ma2005practical} can also be applied to the UBQC protocol. Subsequently, Zhao and Li proposed a blind quantum state preparation protocol with two decoy states~\cite{Zhao2017Blind,Zhao2018Finite,Zhao2020fault}, which further improves the preparation efficiency. Their simulation experiments also show that the preparation protocol with two decoy states is more suitable for long-distance communication.

However, in the preparation of UBQC, the qubits disturbed by noise are prone to errors~\cite{chien2015fault,shan2021multi}. To avoid the accumulation and propagation of these errors in the subsequent computations, quantum error-correcting code~\cite{fujii2015quantum} offers a viable solution to correct error qubits in the preparation process. Tan Xiaoqing et al.~\cite{tan2021fault} proposed a fault-tolerant framework for blind quantum computing, which used 7-qubit CSS code to encode the logical GHZ states to overcome the collective-dephasing noise and the collective-rotating noise. However, the client was required to have the ability of single-qubit measurement on the third qubit of the logical GHZ state and share the remaining Bell state with the server. Morimae and Koshiba~\cite{morimae2019impossibility} have shown that it is impossible for the classical client to implement perfectly secure one-round delegated quantum computing. Therefore, the client at least requires access to the quantum channel or other quantum properties in a realistic situation. In order to reduce the client's burden and dependence on quantum, we delegate quantum error-correcting code preparation to Bob for implementation. Since the delegated preparation is implemented in the measurement-based quantum computation (MBQC) framework~\cite{raussendorf2003measurement,broadbent2010measurement}, the encoding circuit is required to be converted into graph states to prepare the quantum error-correcting codes in the MBQC manner. 

Combined with two-decoy-state RBSP, a remote quantum error-correcting code preparation protocol on cluster state is presented to correct errors in blind quantum computation. In our protocol, the cluster states are used as graph states to prepare quantum error-correcting codes that are unknown to Bob. Then, these codes can be considered as encoded logical qubits for subsequent fault-tolerant blind quantum computation. The security of the proposed protocol is proven theoretically to be $\epsilon$-blind. Finally, the lower bound of quantum resource consumption is estimated, i.e., the number of required pulses. When the prepared quantum error-correcting codes have the same qubit error rate, our protocol can reduce the number of pulses, which will contribute to the practical application of UBQC in \mbox{the future.}

The remainder of the paper is structured as follows: In Section \ref{sec-BQC}, we introduce the basic knowledge. In Section \ref{sec3}, we present a quantum error-correcting code preparation protocol on cluster state to correct qubit errors. Furthermore, we demonstrate that the protocol is $\epsilon$-blind and estimate the quantum resource consumption. In \mbox{Section \ref{sec4}}, simulation results show the quantum resource consumption of coding and non-coding protocols in the case of the same qubit error rate. In Section \ref{sec5}, the necessary conclusions are drawn.

\section{Technical Preliminaries}
\label{sec-BQC}
\subsection{Qubit Error}
Due to environmental noise, it is inevitable that errors occur in quantum computations. 
Therefore, we need to build a general error model to correct qubit errors. In quantum Hilbert space, a quantum state $|\psi\rangle$ can be described as 
\vspace{-3pt}
\begin{eqnarray}\label{eq-psi}
|\psi\rangle=\alpha|0\rangle+\beta|1\rangle,
\end{eqnarray}
where$\left| 0 \right\rangle  = {\left( {\begin{array}{*{20}{c}}1&0\end{array}} \right)^T},\left| 1 \right\rangle  = {\left( {\begin{array}{*{20}{c}}0&1\end{array}} \right)^T}$ are the basis vectors in the Hilbert space, and $|\alpha|^2+|\beta|^2=1$. 

After the noise has occurred, the evolution of the quantum state $|\psi\rangle$ contains four cases: (1) no error, (2) the bit error $\left|0\right\rangle\leftrightarrow \left|1\right\rangle$, (3) the phase error $\alpha|0\rangle+\beta|1\rangle \leftrightarrow \alpha|0\rangle-\beta|1\rangle$, and (4) both bit and phase errors. Hence, an error operator $E_{i}$ can be considered a linear combination of identity operator $I$, bit flip operator $X$, phase flip operator $Z$, and bit-phase flip operator $Y$~\cite{preskill1998fault}. These Pauli operators are
\vspace{-3pt}
\begin{eqnarray}\label{eq-pauli-operator}
I = \left( {\begin{array}{*{20}{c}}
1&0\\
0&1
\end{array}} \right),X = \left( {\begin{array}{*{20}{c}}
0&1\\
1&0
\end{array}} \right),Y = \left( {\begin{array}{*{20}{c}}
0&{ - i}\\
i&0
\end{array}} \right),Z = \left( {\begin{array}{*{20}{c}}
1&0\\
0&{ - 1}
\end{array}} \right).
\end{eqnarray}
To simplify the analysis, suppose the noise occurs on a single qubit, so the quantum state $E_{i}|\psi\rangle$ can be described as a superposition state of the four possible states $I|\psi\rangle$, $X|\psi\rangle$, $Z|\psi\rangle$, and $Y|\psi\rangle$~\cite{nielsen2002quantum}. If one measures the error quantum state $E_{i}|\psi\rangle$, it collapses into one of four states. In order to correct the qubit error, we need to perform a diagnostic process to determine which error occurred in four possible cases and then utilize Pauli operators to act on the error qubit to correct it.
\subsection{Quantum Error-Correcting Codes}
In the error-correcting process, the error syndrome measurement is used to diagnose which qubit is in error. With the increasing number of qubits in code, it is more difficult to determine where the error occurred. As a result, we use seven qubits to encode one qubit, which can correct a qubit error in an encoded block, which is called the 7-qubit Steane code,$[[7,1,3]]$. The encoded logical qubit base $\{|0\rangle_{L},|1\rangle_{L}\}$ is shown as follows:\vspace{-3pt}
\begin{eqnarray}\label{eq-logical-qubits}
\begin{aligned}
{\left| 0 \right\rangle _L} &= \frac{1}{{2\sqrt 2 }}\left( {\left| {0000000} \right\rangle  + \left| {0001111} \right\rangle  + \left| {0110011} \right\rangle  + \left| {0111100} \right\rangle } \right.\\
& + \left. {\left| {1010101} \right\rangle  + \left| {1011010} \right\rangle  + \left| {1100110} \right\rangle  + \left| {1101001} \right\rangle } \right)\\
{\left| 1 \right\rangle _L} &= \frac{1}{{2\sqrt 2 }}\left( {\left| {1111111} \right\rangle  + \left| {1110000} \right\rangle  + \left| {1001100} \right\rangle  + \left| {1000011} \right\rangle } \right.\\
& + \left. {\left| {0101010} \right\rangle  + \left| {0100101} \right\rangle  + \left| {0011001} \right\rangle  + \left| {0010110} \right\rangle } \right)
\end{aligned}
\end{eqnarray}

\begin{figure}[H]
	\includegraphics[width=0.75\textwidth]{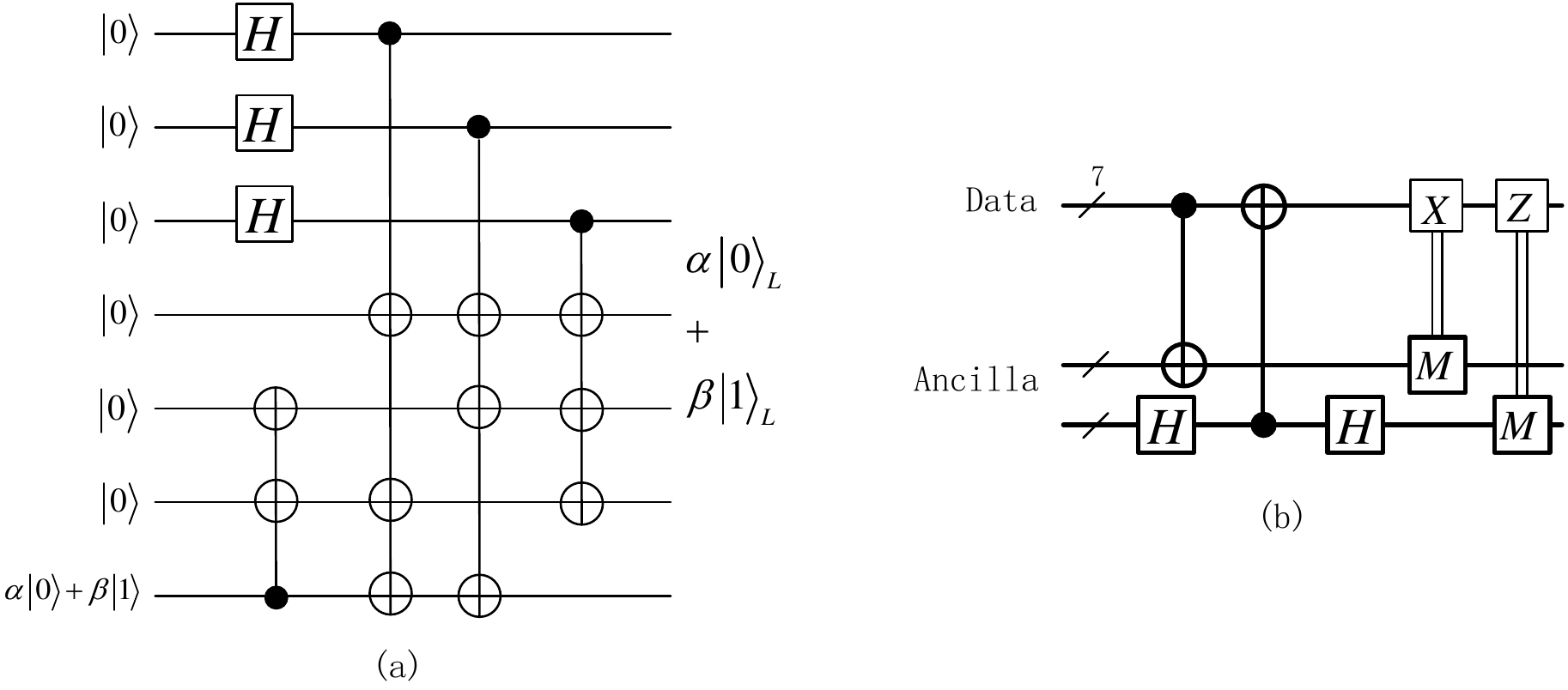}
	\caption{\label{fig-encode-circuit}The quantum encoded and syndrome measurement circuits of the Steane code [[7,1,3]]. (\textbf{a}) A 7-qubit encoded logical qubit is prepared by a data qubit and six ancilla qubits. (\textbf{b}) The bit and phase errors occurring in the encoded block can be diagnosed with two Steane ancilla states and recovered by Pauli gates.}
\end{figure}

Based on the encoding principle of Steane code~\cite{preskill1998fault}, we can design the encoding circuit of quantum error-correcting code, as shown in Figure~\ref{fig-encode-circuit}a. The unknown data qubit and six ancilla qubits can be used to encode an encoded logical qubit. According to the error syndrome measurement circuit in Figure~\ref{fig-encode-circuit}b, one uses two Steane ancilla states to diagnose the bit and phase error syndromes, respectively. Finally, the Pauli gates are used to act on the error qubits to correct them in each encoded block.
\vspace{-6pt}

\subsection{Realization of Quantum Gates on Cluster State}

For a quantum circuit, we can use the ordered quantum gates to describe the computational process. Any quantum gate can be transformed into a combination of quantum gates in the universal gate group, such as $\{CNOT, H, \pi/8\}$. Hence, only if we can realize every quantum gate in the universal logic gate group can we implement arbitrary \mbox{quantum computing.}

\begin{figure}[H]
	\includegraphics[width=0.75\textwidth]{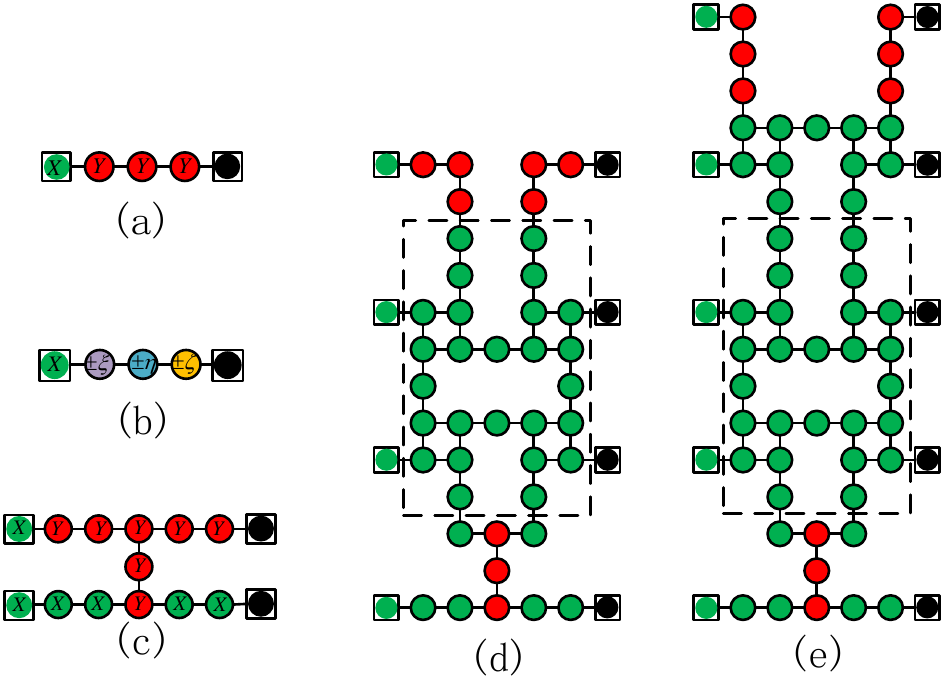}
	\caption{\label{fig-gate-cluster}Implementation of basic quantum gates on cluster state. Each circle represents a qubit, and the left-most squares represent the input qubits, the right-most squares denote the output qubits. Each green circle represents a cluster qubit measured in the eigenbasis of Pauli gate X, the red denotes Y. The $\xi$, $\eta$, and $\zeta$ represent the measurement angles $\delta$ of the measurement basis $M(\delta)$.   (\textbf{a}) The Hadamard gate. \mbox{(\textbf{b}) A} general rotation gate. (\textbf{c}) The CNOT gate applied between adjacent logical qubits. (\textbf{d}) The CNOT gate applied between two logical qubits separated by an even number. (\textbf{e}) The CNOT gate applied between two logical qubits separated by an odd number.}
\end{figure}

In the framework of MBQC~\cite{raussendorf2001one}, we know that quantum gates can be realized by multi-particle entangled graph states, such as cluster state and brickwork state~\cite{broadbent2009universal,raussendorf2003measurement}. Note that the $[[7,1,3]]$ encoded circuit consists of Hadamard gates and CNOT gates in Figure~\ref{fig-encode-circuit}. Hence, we only need to realize these two types of quantum gates on graph states. Since quantum gates in brickwork state only act on the neighboring qubit, the CNOT gate acting on non-neighboring qubits needs many SWAP gates to be realized, which is very inefficient. Hence, we use the cluster state to realize these Hadamard and CNOT gates in the encoding circuit, as shown in Figure \ref{fig-gate-cluster}.
\vspace{-6pt}

In Figure \ref{fig-gate-cluster}, these diagrams are used to illustrate the realization of quantum gates on cluster states. (a) represents the pattern to realize the Hadamard gate. Each circle symbolizes a qubit, and the lines indicate entangling operators. Controlled-Z (CZ) gates are employed to interact between neighboring qubits, preparing the entangled state. The qubits are measured in a specifically selected basis, with the green (red) circle representing a measurement in the eigenstates of the $X$ ($Y$) Pauli gate. The measurement basis $M(\delta)$ is determined by the measurement angle $\delta$. The eigenbasis of $X$ ($Y$) corresponds to \mbox{$\delta=0 (\pi/2)$.} The computational basis $\{|0\rangle,|1\rangle\}$ is the eigenbasis of $Z$. (b) illustrates a general one-qubit rotating quantum gate through one-qubit measurement in a cluster state, where the measurement basic angles $\pm \xi,\pm \eta,\pm \zeta$ depend on the measurement results of other qubits. 
(c), (d), and (e) show the CNOT gate applied between different distance logical qubits. One can repeat the rectangle parts enclosed by the black dashed line in (d) and (e) to deal with any separation. In quantum computing in a cluster state, the $\{|0\rangle,|1\rangle\}$ basis is used to eliminate the redundant qubits, and the adaptive basis $M(\theta)$ is used to measure the remaining qubits to implement any quantum gate.

\section{Quantum Error-Correcting Code Preparation on Cluster State}\label{sec3}
In the field of quantum computation, it is widely recognized that cluster states can be used to realize arbitrary quantum gates. Indeed, the essence of a quantum circuit resides in its meticulously ordered sequence of quantum gates, forming the fundamental backbone of quantum computing. Consequently, each quantum circuit can be efficiently performed in a cluster state~\cite{raussendorf2003measurement}. By harnessing the power of cluster states, the quantum encoding circuit can also be feasible, enabling the generation of quantum error-correcting codes. These encoded qubits can then be utilized to construct novel brickwork states to facilitate the realization of fault-tolerant quantum computing. 

\begin{figure}[H]
	\includegraphics[width=0.95\textwidth]{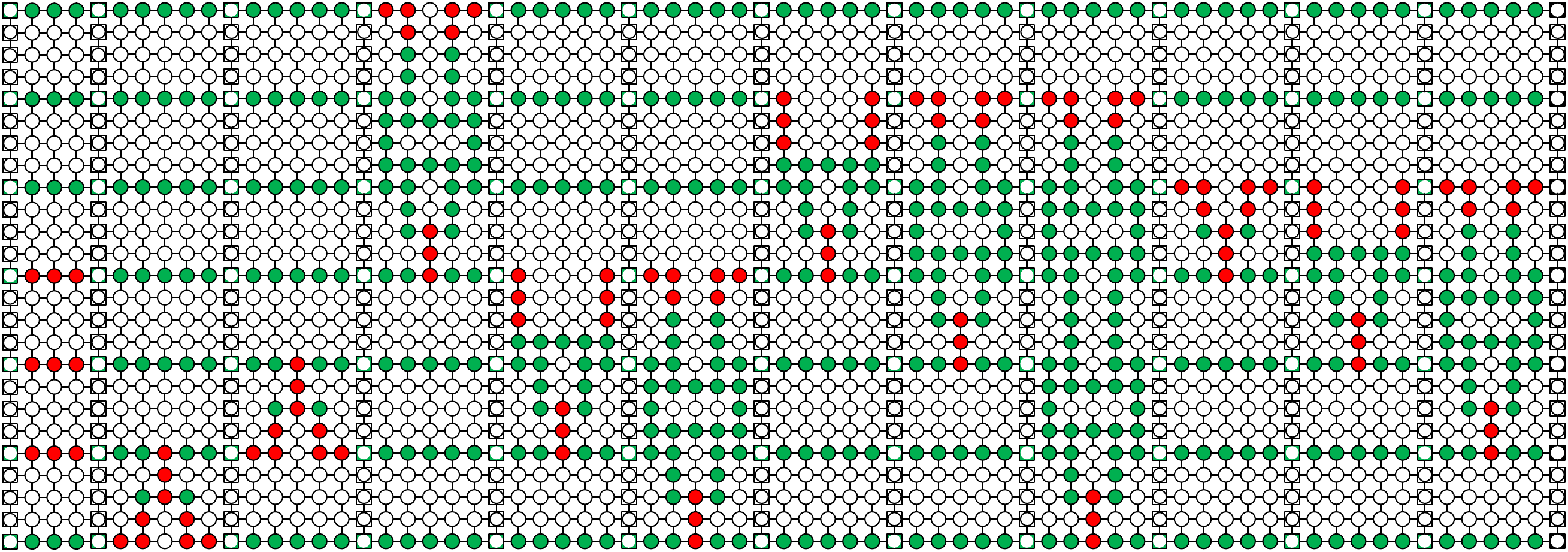}
	\caption{The implementation of the [[7,1,3]] encoding circuit on cluster state. The measured cluster qubits are denoted by the green, red, and white circles, representing the measurements in the eigenbases of $X$, $Y$, and $Z$, respectively.}
	\label{fig-encoding-cluster}
\end{figure}

According to the implementation of quantum gates on cluster state in the preliminaries, the encoding circuit of [[7,1,3]] code needs to be transformed into MBQC on cluster state, then the quantum error-correcting code preparation can be realized through measuring each qubit on cluster state, as shown in Figure \ref{fig-encoding-cluster}. In the preparation process, Bob initiates the construction of the initial cluster state, followed by the judicious utilization of the computational basis $\{|0\rangle,|1\rangle\}$ to eliminate the redundant qubits based on the precise positional information provided by Alice. In accordance with Alice's designated measurement bases $M(\delta)$, the remaining qubits of the cluster state serve as a platform for preparing quantum error-correcting codes. These bases are meticulously designed to be orthogonally projected onto the states $|\pm_{\delta}\rangle=(|0\rangle \pm e^{i\delta}|1\rangle)/\sqrt{2}$, where $\delta \in [0,2\pi]$ designates the measurement angle. Note that the specific values of $\delta$ such as $\delta=0$ or $\pi/2$ corresponds to the $X$ or $Y$ Pauli measurement, respectively. It is important to emphasize that these measurements are inherently destructive. Obviously, the measurement is understood as a destructive measurement. The measurement outcome of qubit $i$ is denoted by $s_{i} \in \mathbb{Z}_2$. Our convention dictates that $s_{i}=0$ when the state collapses to $|+_{\delta}\rangle$ as a result of the measurement, and $s_{i}=1$ when it collapses to $|-_{\delta}\rangle$.
\vspace{-6pt}

During the process of removing redundant qubits, an inherent challenge arises concerning the potential leakage of quantum gates used, which could inadvertently disclose the underlying structural information of the cluster state to Bob. However, Alice's primary concern lies in ensuring the security of the encoded logical qubit $|+_{\theta_i}\rangle_{L}$ during the preparation, while the quantum gates employed in the encoded circuit can be made public to Bob. To accomplish preparation, three distinct measurement bases are employed, including $\{|0\rangle,|1\rangle\}$, $M(0)$ and $M(\pi/2)$ corresponding to the eigenbases of the Pauli gates $Z, X, Y$, respectively. These measurement bases exhibit a notable characteristic: their independence from the polarization angle $\theta_i$. As a result, the information pertaining to $\theta_i$ remains entirely secure and undisclosed to Bob throughout the process. These findings shed light on the potential of cluster states for secure and efficient quantum information processing in quantum error-correcting code preparation.


In UBQC, the delegated preparation process contains Alice's preparation, Bob's preparation, and the interaction measurement. In Alice's preparation, Alice sends $N$ data pulses to Bob, which comprise signal states and two decoy states, and their polarization is ${\rho ^\sigma }$, where $\sigma$ is chosen randomly from $\left\{ {\frac{{k\pi }}{4}:0 \le k \le 7} \right\}$. In addition, Alice needs to send a series of ancilla pulses to Bob, and their polarization is $|+\rangle$, which is allowed to be public. In Bob's preparation, the required qubit $|+_{\theta_{i}}\rangle$ can be generated based on non-coding RBSP~\cite{Zhao2018Finite}. Among the generated qubit $|+_{\theta_{i}}\rangle$ and a series of ancilla qubits, Bob utilizes CZ gates to entangle them to build an initial cluster state. During the interaction measurement, Bob follows Alice's instructions and eliminates the redundant qubits from the initial cluster state in the computational basis, as shown in the white circles in Figure \ref{fig-encoding-cluster}. According to Alice's measurement basis $M(\delta_{ij}),\delta_{ij}\in\{0,\pi/2\}$, the remaining qubits on cluster state are measured sequentially until the final outcome is obtained. Finally, Bob can achieve the required quantum error-correcting codes. As depicted in Protocol \ref{protocol1-cluster}, we propose an innovative protocol for remote quantum error-correcting code preparation on cluster state, which lays the groundwork for subsequent fault-tolerant blind quantum computing. 
 
\vspace{6pt}
\begin{algorithm}[H]
	\caption{A remote quantum error-correcting code preparation on cluster state}
	\label{protocol1-cluster}
	\KwIn {data pulses with polarization $\rho^{\sigma}$, $\sigma \in_{R} \{ k\pi /4:\;0 \le k \le 7\}$ including signal and two decoy states; ancilla pulses with polarization $|+\rangle$.}
	\KwOut {the encoding logical qubits: $\left\{ {|{ + _{{\theta _i}}}\rangle_{L} } \right\}_{1}^S$.}
	Alice uses the laser to send the required data and ancilla pulses to Bob.
	
	\For{$i=1$ to $S$;}
	{
		Bob utilizes non-coding RBSP to prepare qubit $|+_{\theta_i}\rangle$, then uses it and a series of ancilla qubits $\{|+\rangle\}$ to build the cluster state.
		
		\For{$x=1$ to $m$, $y=1$ to $n$;}
		{
			Based on Figure \ref{fig-encoding-cluster}, Bob performs the [[7,1,3]] encoding circuit on cluster state
			
			\lIf{$q_{xy}$ is white}{Bob performs measurement in $\{|0\rangle,|1\rangle\}$ }
			
			\lIf{$q_{xy}$ is green}{Bob performs measurement in $M(0)$}
			
			\lIf{$q_{xy}$ is red}{Bob performs measurement in $M(\pi/2)$}			
		}
		Bob generates quantum error-correcting code $\left\{ {|{ + _{{\theta _i}}}\rangle_{L} } \right\}$.
	}
	\Return  $\left\{ {|{ + _{{\theta _i}}}\rangle_{L} } \right\}_{1}^S$.
\end{algorithm}

\vspace{9pt}

The [[7,1,3]] encoding circuit is designed based on the generator matrix, and its correctness had been proven by Preskill~\cite{preskill1998fault}. Since the preparation on cluster state is equivalent to the encoding circuit model, our protocol can generate the correct quantum error-correcting codes. In order to ensure Alice's information privacy, the encoding logical qubits (quantum error-correcting codes) are required to be unknown to Bob in UBQC. In other words, the polarization angle $\theta$ is always unknown to Bob in the preparation process of the encoding logical qubits $|{ +_{\theta}}\rangle_{L}$. 

In UBQC, the client and server share a joint state that describes the evolution of the entire system. The ideal state is symbolized as $\pi_{AB}^{ideal}$ that effectively guarantees the client's information against potential malicious attacks. During the preparation, the ideal state $\pi_{AB}^{ideal}$ evolves into a set of states, which can be conveniently represented by $\mathcal{F}(\pi_{AB}^{ideal})$. To ensure robust security against any server actions, it is imperative that all states with this set remain equally blind. To assess the security level achievable in practical implementation, we explore scenarios where the client transmits a realistic state $\rho^{\theta_i}$ to replace a perfect state $|+_{\theta_{i}}\rangle$. In the following, we introduce the concept of almost blindness by quantifying the proximity between the realistic joint state and the ideal joint state, i.e., $\epsilon$-blind 
 \cite{dunjko2012blind}.

\begin{Definition}\label{def-epsilon}
A UBQC protocol with imperfect preparation is $\epsilon$-blind if the trace distance between the ideal joint state $\pi_{AB}^{\theta_i}$ and realistic joint state $\pi_{AB}^{\rho^{\theta_i}}$ is less than $\epsilon>0$. Finally, we have:
\vspace{-3pt}
\begin{equation}\label{eq-definition-blind}
\mathop {\min }\limits_{\pi _{AB}^{\theta_{i}} \in \mathcal{F}} \frac{1}{2}\left\| {\pi _{AB}^{\{ {\rho ^{\theta _i}}\} } - \pi _{AB}^{\left\{ {{\theta _i}} \right\}}} \right\| \le \varepsilon.
\end{equation}
\end{Definition}

\begin{Theorem}\label{thm-preblind}
Our preparation Protocol \ref{protocol1-cluster} is ${\epsilon}$-blind to Bob.
\end{Theorem}

\begin{proof}
	
In our protocol, the measurement-based quantum computation is used to realize quantum error-correcting code preparation on cluster state. The prepared data qubit $|+_{\theta}\rangle$ and a series of ancilla qubits $|+\rangle$ can be considered as quantum resources to build the cluster state. Based on the RBSP protocol \cite{dunjko2012blind,Zhao2018Finite}, the generated data qubit $|+_{\theta}\rangle$ is ${\epsilon}$-blind. In order to demonstrate the ${\epsilon}$-blindness of the encoded logical qubit $|+_{\theta}\rangle_{L}$, we only need to prove that the encoding circuit on cluster state is ${\epsilon}$-blind. In other words, the angle of polarization $\theta$ of $|+_{\theta}\rangle_{L}$ in the preparation process is unknown to Bob.

To be more precise, it is necessary to illustrate that the measurements on cluster state do not reveal any information about polarization angle $\theta$. The cluster state is built by multiple entanglement qubits. For the purpose of simplicity, it suffices to prove that the polarization angle $\theta$ is ${\epsilon}$-blind to Bob during the measurements on the minimum cluster state. Further, we analyze three kinds of measurements on cluster state according to the [[7,1,3]] encoding circuit. The eigenstates of $X,Y,Z$ are used to measure the qubits on the minimum cluster state, respectively. We explore whether there is information leakage of the polarization angle $\theta$ during the measurements. 
\begin{eqnarray}\label{eq-Mxy-cluster}
\begin{aligned}
{\left| {{ + _\theta }} \right\rangle _1} \otimes {\left|  +  \right\rangle _2} &= \frac{1}{{\sqrt 2 }}{\left( {\left| 0 \right\rangle  + {e^{i\theta }}\left| 1 \right\rangle } \right)_1} \otimes {\left|  +  \right\rangle _2}\\
&\xrightarrow{CZ} \frac{1}{{\sqrt 2 }}{\left| 0 \right\rangle _1} \otimes {\left|  +  \right\rangle _2} + \frac{{{e^{i\theta }}}}{{\sqrt 2 }}{\left| 1 \right\rangle _1} \otimes {\left|  -  \right\rangle _2}\\
&\xlongequal[]{M(X)} \frac{{{{\left|  +  \right\rangle }_1} \otimes \left( {H{{\left| {{ + _\theta }} \right\rangle }_2}} \right) + {{\left|  -  \right\rangle }_1} \otimes \left( {X \cdot H{{\left| {{ + _\theta }} \right\rangle }_2}} \right)}}{{\sqrt 2 }}\\
&\xlongequal[]{M(Y)}\frac{{{{\left| {{ + _{\pi /2}}} \right\rangle }_1} \otimes \left( {XHS\left| {{ + _\theta }} \right\rangle_{2} } \right) + {{\left| {{ - _{\pi /2}}} \right\rangle }_1} \otimes \left( {HS\left| {{ + _\theta }} \right\rangle_{2} } \right)}}{{\sqrt 2 }}
\end{aligned}
\end{eqnarray}

The first qubit $|+_{\theta}\rangle$ and the second qubit $|+\rangle$ are used to build the minimum cluster state using CZ gate, as shown in Figure \ref{fig-min-cluster}. If the first qubit is measured in the eigenstates of $X$ gate, as shown in Figure \ref{fig-min-cluster}(a), the quantum state is $|+\rangle$ or $|-\rangle$ with an equal probability. The second qubit is evolved into $(X)^{k} \cdot H|+_{\theta}\rangle_{2},k\in\{0,1\}$, where $k$ represents the measurement result of the first qubit, as shown in Equation \eqref{eq-Mxy-cluster}. If the eigenstates of $Y$ gate are used to measure the first qubit, as shown in Figure \ref{fig-min-cluster}(b), the polarization state of the second qubit is evolved into $(X)^{k}\cdot H\cdot S|+_{\theta}\rangle_{2},k\in\{0,1\}$, $k$ is also determined by measurement result of the first qubit, as shown in Equation \eqref{eq-Mxy-cluster}. It can be seen that the information of the first qubit is transmitted to the second qubit through MBQC without loss of information. If the first qubit $|+\rangle$ and the second qubit $|+_{\theta}\rangle$ are entangled into the minimum using CZ gate, as shown in Figure \ref{fig-min-cluster}(c), and the eigenstates of $Z$ gate are used to measure the first qubit, the second qubit is still $|+_{\theta}\rangle$, as shown in Equation \eqref{eq-Mz-cluster}. Obviously, the remaining qubits are unchanged when the redundant qubits are eliminated by $Z$. 
\vspace{-3pt}
\begin{eqnarray}\label{eq-Mz-cluster}
\begin{aligned}
{\left|  +  \right\rangle _1} \otimes {\left| {{ + _\theta }} \right\rangle _2} &= \frac{1}{{\sqrt 2 }}{\left( {\left| 0 \right\rangle  + \left| 1 \right\rangle } \right)_1} \otimes {\left| {{ + _\theta }} \right\rangle _2}\\
&\xrightarrow{CZ}\frac{1}{{\sqrt 2 }}{\left| 0 \right\rangle _1} \otimes {\left| {{ + _\theta }} \right\rangle _2} + \frac{1}{{\sqrt 2 }}{\left| 1 \right\rangle _1} \otimes {\left| {{ - _\theta }} \right\rangle _2}\\
&\xlongequal[]{M(Z)}\frac{{{{\left| 0 \right\rangle }_1} \otimes {{\left| {{ + _\theta }} \right\rangle }_2} + {{\left| 1 \right\rangle }_1} \otimes \left( {Z{{\left| {{ + _\theta }} \right\rangle }_2}} \right)}}{{\sqrt 2 }}
\end{aligned}
\end{eqnarray}
\vspace{-6pt}
\begin{figure}[H]
	\includegraphics[width=0.75\textwidth]{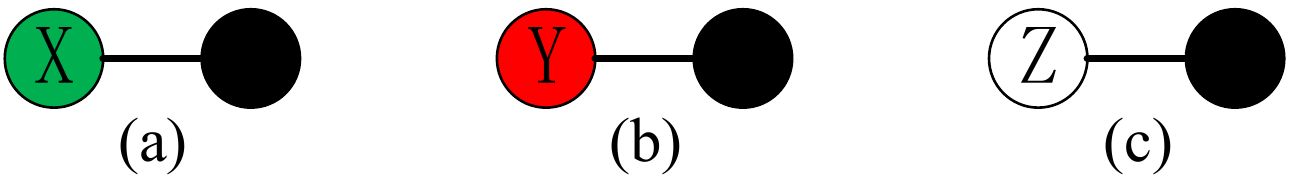}
	\caption{\highlight{The } 
cluster state with two qubits. The green, red, and white represent the measurements of $X$, $Y$, and $Z$ gates. (a) The first qubit of the cluster state is measured in the eigenbasis of $X$. (b) The first qubit is measured in Y. (c) The first qubit is measured in Z.}
	\label{fig-min-cluster}
\end{figure}

Thus, the polarization angle $\theta$ is irrelevant to the eigenstates of $X, Y$, and $Z$. During the measurements on cluster state, Bob can not obtain any information about the angle of polarization $\theta$. Hence, the generated encoded logical qubit $|+_{\theta}\rangle_{L}$ is ${\epsilon}$-blind in our Protocol \ref{protocol1-cluster} if the input data qubit $|+_{\theta}\rangle$ is ${\epsilon}$-blind.
\end{proof}

In our protocol, these pulses sent by Alice consist of two parts, data and ancilla. Based on non-coding RBSP, note that data pulses are employed to generate the required data qubits $\{|+_{\theta_i}\rangle\}_1^S$. Each data qubit is then transformed into its corresponding logical qubit as required. Consequently, the number of data pulses used aligns with that of the non-coding RBSP, denoted as $N^{d} $~\cite{Zhao2020fault}. The number of ancilla pulses depends on the cluster state used to prepare encoded states, as shown in Figure \ref{fig-encoding-cluster}. It is important to highlight that the number of ancilla qubits is a constant if one encoded qubit is prepared, denoted as $C$. The transmittance of the quantum channel between client and server directly affects the probability of the pulses being received, denoted as $T$, and the scale of quantum computation is denoted as $S$, the number of required ancilla pulses $N^{a}=CS/T$. Hence, we can derive the lower bound of the total number of pulses in Protocol \ref{protocol1-cluster}.
\vspace{-6pt}
\begin{eqnarray}\label{eq-N-QEC}
\begin{aligned}
N &= {N^d} + {N^a}\\
&\ge {N^{L,{v_1},{v_2}}} + \frac{{CS}}{T}\\
&=\frac{S}{T}\left[ {\frac{{\ln \left( {\epsilon/S} \right)}}{{{p_\mu }\mu \ln \left( {1 - p_1^{L,{v_1},{v_2}}} \right)}} + {\rm{C}}} \right]
\end{aligned}
\end{eqnarray}
where $\epsilon$ represents the secure parameter. $\mu$,$v_1,v_2$ are the average photon number of signal states and two decoy states, respectively. $p_1$ represents the proportion of single-photon states in the signal states, and the lower bound is denoted as ${p_1^{L,v_1,v_2}}$~\cite{Zhao2018Finite}. The probabilities of signal pulses chosen by the client are defined as $p_{\mu}$. $S$ is the computational scale, and $T$ represents the transmittance of the quantum channel.

We assume that the qubit errors are independent. We can repeatedly use ancilla qubits in the correction, so we do not consider the ancilla qubit consumption in the correction process, as shown in Figure~\ref{fig-encode-circuit}b. In the $[[7,1,3]]$ code, we note that each encoded block can correct one error. Therefore, before encoding, the error probability of each qubit is $e$ ($e<1$), and after encoding, it changes to $e^{2}$. In a quantum computation with computation scale $S$, the error rate of each generated qubit based on non-coding RBSP~\cite{Zhao2018Finite} is $e$. If an error qubit occurs in all generated qubits, the preparation process fails. Since the error rate of each encoded qubit is $e^2$ in the preparation of Protocol \ref{protocol1-cluster}, the successful probability of encoded preparation is derived as $(1-e^2)^S$. We consider the non-coding RBSP protocol as a repeated Bernoulli experiment, then the success probability of preparation is $(1-e)^{S}$. By repeating the process $k$ times, the success probability can be calculated as $1-[1-(1-e)^{S}]^{k}$. In the scenario where the success probabilities are equal for both coded and non-coded cases, specifically indicated by $1-[1-(1-e)^{S}]^{k}=(1-e^2)^S$, the number of repetitions, denoted as $k$, can be determined using the following derivation:
\begin{equation}\label{eq-repeat-k}
k = \ln [1 - {(1 - {e^2})^S}]/\ln [1 - {(1 - e)^S}].
\end{equation}
Hence, when the generated qubit error rate in the coding case is $e^2$, the non-coding RBSP needs to repeatedly send $kN^{d}$ pulses to achieve the same generated qubit error rate. 

In UBQC, we assume that Alice has a laser transmitter with frequency $f$, and Bob has a full-fledged quantum computer. By deriving the preparation efficiency of UBQC, which represents the rate at which qubits are generated per second, we can gain valuable insights into the system's performance. According to Equation \eqref{eq-N-QEC}, the upper bound of the efficiency for the concatenation code can be estimated as
\vspace{-3pt}
\begin{equation}\label{eq-efficiency}
\begin{aligned}
E &= S \cdot f/{N}\\
&\le S \cdot f/\left( {{N^{L,{v_1},{v_2}}} + N^a} \right).
\end{aligned} 
\end{equation}

In summary, our Protocol \ref{protocol1-cluster} can prepare ${\epsilon}$-blind quantum error-correcting codes to correct qubit errors. Under the condition of the same generated qubit error rate, our approach can reduce the number of weak coherent pulses and improve preparation efficiency compared with the non-coding RBSP protocol. 

\section{Simulation Results}\label{sec4}
An Intel(R) Core(TM) i7-6700HQ CPU, 12.0GB RAM, and Win 10 pro OS were utilized to implement the simulations through MATLAB software. The transmittance $T$ contains the outside fiber, the inside fiber, and the detection efficiency between Alice and Bob, which can be calculated as follows:
\vspace{-3pt}
\begin{equation}
T = {t_s}\cdot{\eta _s}\cdot{10^{ - \alpha L/10}}
\end{equation}
where $\alpha$ represents the loss coefficient in decibels per kilometer (dB/km), $L$ represents the length of the fiber optic cable in kilometers (km), $t_{s}$ stands for the internal transmittance of Bob's optical components, and ${\eta_{s}}$ represents the detector efficiency. The average photon number of signal states is denoted as $\mu$, and two decoy states are represented as $v_1, v_{2}$. The signal state is chosen with probability $p_{\mu}$, and two decoy states are chosen with $p_{v_1}, p_{v_2}$. The scale of quantum computation is $S$. The secure parameter is symbolized as $\epsilon$. The error rate per qubit is denoted as $e$. According to Figures \ref{fig-encode-circuit} and \ref{fig-encoding-cluster}, the required number of ancilla qubits can be determined, i.e., $C=1774$. The repetition frequency of the laser transmitter is $f=1$ MHZ. The relative parameters setting in our simulation are shown in Table \ref{table-parameter} (refer to the data in \cite{Xu2015Blind,ma2005practical}).

\begin{table}[H]\setlength{\tabcolsep}{2.9mm}
	\caption{The simulation parameters for our protocol.}
	\label{table-parameter}
	\begin{tabular}{ccccccccccccc}
		\toprule
		\boldmath{$\alpha$} & \boldmath{$t_{S}$} & \boldmath{$\eta_{S}$}  & \boldmath{$\mu$} & \boldmath{$v_{1}$} & \boldmath{$v_{2}$} & \boldmath{$p_{\mu}$} & \boldmath{$p_{v_1}$} & \boldmath{$p_{v_2}$} & \boldmath{$S$} & \boldmath{$\epsilon$} & \boldmath{$e$}\\
		\midrule
		0.2 & 0.45 & 0.1 & 0.6 & 0.125 & 0 & 0.9 & 0.05 & 0.05 &  1000 & $10^{-10}$ & $0.01$\\
		\bottomrule
	\end{tabular}
\end{table}

\begin{figure}[H]
	\includegraphics[width = 0.65\textwidth]{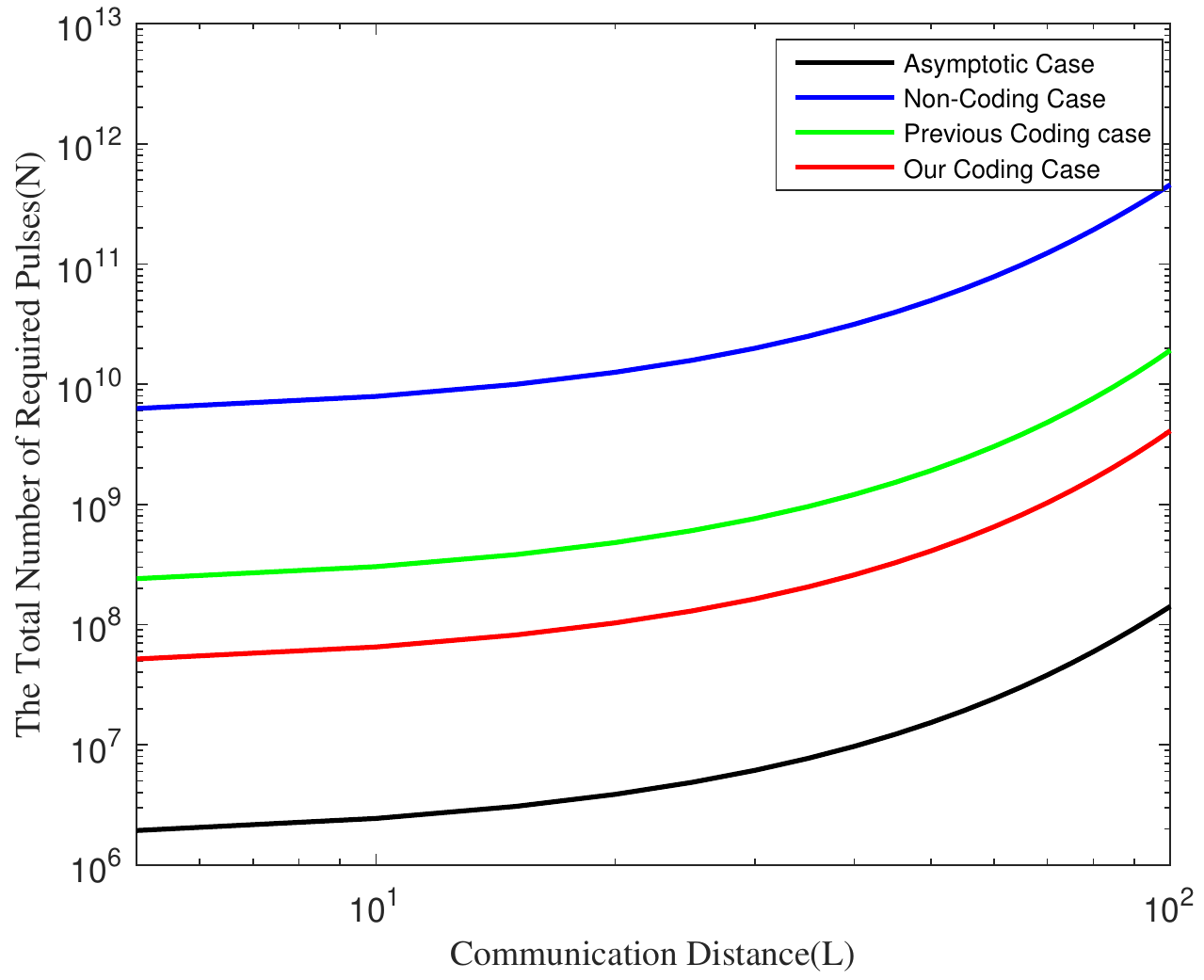}
	\caption{The relationship between the total number of pulses $N$ case and the communication distance $L$ with the same qubit error rate. The red line represents the quantum resource consumption of Protocol \ref{protocol1-cluster} with coding, the green line represents the simulation results of the previous coding case~\cite{Zhao2020fault}, and the blue line depicts the results in the non-coding case~\cite{Zhao2018Finite}. The simulation results depicted by the black line show the outcomes in the asymptotic scenario, characterized by an infinitely large data size and near-perfect preparation of qubits.}
	\label{fig-one-Concatenation-LvsN}
\end{figure}

In Figure~\ref{fig-one-Concatenation-LvsN}, note that the number of pulses in the coding case is comparatively lower than that in the non-coding RBSP~\cite{Zhao2018Finite} with the same qubit error rate and is closer to the asymptotic scenario (with infinite data-size and no qubit error~\cite{ma2005practical,Zhao2018Finite}). The error rate of the generated qubits using Protocol \ref{protocol1-cluster} is significantly lower than that of the non-coding RBSP. In order to obtain the same error rate, the non-coding case is required to repeat $k$ times. Under the encoding condition of the same qubit error rate, the number of pulses required in our Protocol \ref{protocol1-cluster} is lower than that of the previous coding case~\cite{Zhao2020fault}. On the one hand, since our approach uses the cluster state instead of the brickwork state to realize remote delegated preparation, it can reduce many SWAP gates when utilizing CONT gates between non-adjacent qubits. On the other hand, our approach requires the CNOT gates with a high execution probability to avoid using too many ancilla qubits in the preparation to realize the operations of the fault-tolerant CNOT gates. Both aspects can greatly reduce the consumption of ancilla qubits; thereby, our proposed Protocol \ref{protocol1-cluster} can reduce the number of pulses required.

\begin{figure}[H]
	\includegraphics[width = 0.65\textwidth]{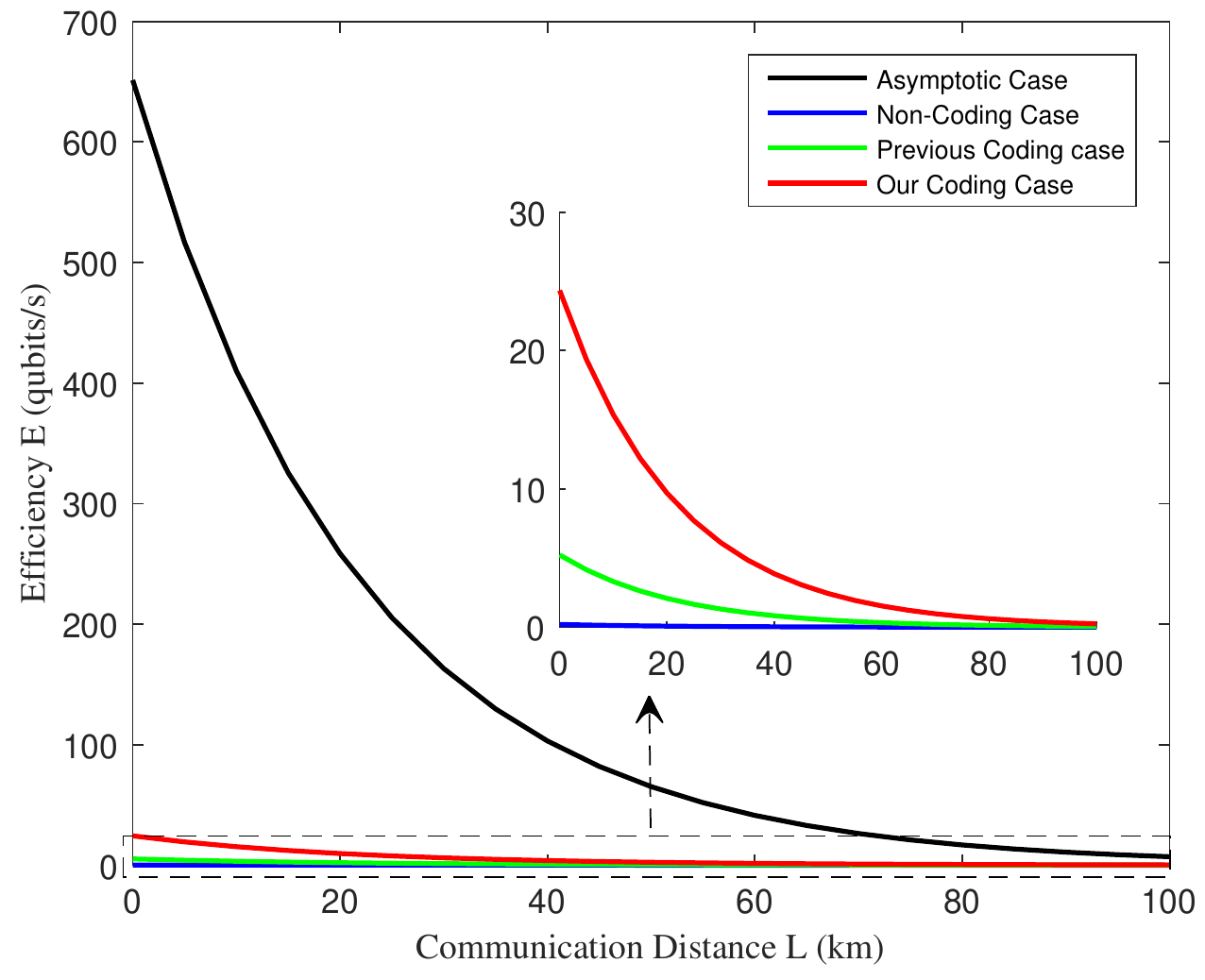}
	\caption{The relationship between preparation efficiency $E$ and the communication distance $L$ with the same qubit error rate. The red line represents the preparation efficiency of Protocol \ref{protocol1-cluster} with coding, the green line represents the simulation results of the previous coding case~\cite{Zhao2020fault}, and the blue line depicts the results in the non-coding case~\cite{Zhao2018Finite}. The simulation results represented by the black line illustrate the outcomes in the asymptotic scenario, characterized by an infinitely large data size and near-perfect preparation of qubits.}
	\label{fig-one-Concatenation-Lvsefficiency}
\end{figure}

In Figure~\ref{fig-one-Concatenation-Lvsefficiency}, the preparation efficiency $E$ in the coding case is less than the non-coding RBSP with the same qubit error rate and is closer to the asymptotic case. Especially, the efficiency of our proposed Protocol \ref{protocol1-cluster} is better than the previous protocols~\cite{Zhao2018Finite,Zhao2020fault} with the same qubit error rate. This reason is that our approach can reduce the consumption of SWAP gates and ancilla qubits. With the increasing distance, the value of efficiency $E$ grows rapidly, implying the significant impact of channel loss and corresponding qubit error rate. 
For long-distance communication, our protocol surpasses the non-coding case, primarily due to the fact that the value of $E$ is closer to the asymptotic scenario (with an infinitely large data size and no qubit error~\cite{ma2005practical,Zhao2018Finite}). These advantages underscore the enhanced performance and efficacy of our protocol, making it a favorable choice for long-distance blind quantum computing.
\vspace{-6pt}
\unskip

Therefore, our protocol holds the potential to enhance not only the preparation efficiency but also the conservation of quantum resources. Furthermore, the quantum error-correcting codes can be used as logical qubits to facilitate subsequent fault-tolerant blind quantum computation. The capability of our protocol ensures an optimized utilization of quantum resources while maintaining the integrity and reliability of the preparation \mbox{in UBQC.}

\section{Conclusions}\label{sec5}
In this study, we propose an innovative protocol for remote quantum error-correcting code preparation on cluster states, which can be used to correct error qubits in the preparation of UBQC. Based on the blindness of the original RBSP protocol, we demonstrate that Protocol \ref{protocol1-cluster} is also $\epsilon$-blind. Alice only needs to send weak coherent pulses, without quantum memory or extensive quantum computing capabilities. By incorporating quantum error-correcting codes, our protocol can reduce the number of pulses when preparing the quantum error-correcting codes with the same qubit error rate. Furthermore, these quantum error-correcting codes prepared by our protocol are unknown logical qubits to Bob, which can be used to build a new encoded brickwork state to realize the fault-tolerant blind quantum computation. Our proposed preparation protocol provides a theoretical basis for the practical application of blind quantum computation in the future. In our approach, when CNOT gates are executed, they need to run with a high probability. This will limit the generalized application of blind quantum computation. Hence, we will continue to research the surface codes and the concatenated codes with high fault tolerance to better balance the quantum resource consumption and the qubit error rate, so as to generalize the preparation protocol of quantum error-correcting codes.

\vspace{6pt}

\authorcontributions{Writing---original draft preparation, Q.Z.; writing---review and editing, Q.Z., H.M., Y.Q., Q.L., and A.A.A.E.-L.; funding acquisition, Q.L.
 All authors have read and agreed to the published version of the manuscript.}

\funding{This work is supported by the National Natural Science Foundation of China (grant Number: 62071151). Also, Ahmed A. Abd El-Latif acknowledges the Talented Young Scientist Program (TYSP) and its support.}

\institutionalreview{This research is not applicable to studies involving humans or animals.}

\informedconsent{This research is not applicable to studies involving humans.}

\dataavailability{In our work, no new real data were created, and all the data were simulated data generated by the parameters in the paper.}

\conflictsofinterest{The authors declare no conflicts of interest.} 



\begin{adjustwidth}{-\extralength}{0cm}

\reftitle{References}

\PublishersNote{}
\end{adjustwidth}
\end{document}